\documentclass{phostproc}
\usepackage{float}

\title{Evolutionary stage of the massive component of the double-lined eclipsing binary V380 Cygni}
\author{Amadeusz Miszuda $^1$,
	Jadwiga Daszy\'nska-Daszkiewicz,
	Wojciech Szewczuk}

\affiliation{Astronomical Institute of the Wroc\l aw University, Kopernika 11, PL-51-622 Wroc\l aw, Poland \\
					$^1$ E-mail: miszuda@astro.uni.wroc.pl}

\shorttitle{ [The evolutionary stage of V380 Cygni}
\shortauthors{A. Miszuda et al.}

\abs{We determine the age of the eclipsing binary V380 Cygni, whose components have a significant mass ratio, $M_B/M_A\approx$ 0.6. 
Because this binary exhibits spectral lines of both components, the values of masses and radii could be determined with a very high accuracy. 
Assuming that both components are on the main sequence, we derive the age of about 16 Myr.
However, in this case the primary needs a large overshooting from the convective core ($f_{\rm{ov}} \approx 0.04$) and higher metallicity ($Z \approx 0.02$). 
It is difficult to tell whether such a high value of overshooting from the convective core is reliable or rather mimics other mixing processes, 
which took place on the whole main sequence evolution, e.g., resulting from diffusion, rotation etc.
If the primary is already in the overall contraction phase or just after, then the age of the system is about 13 Myr.
Based on the available observations, we cannot resolve what is the evolutionary stage of the V380 Cyg system.\\
The analysis of the {\it Kepler} light curve does not show any significant periodicity besides the orbital frequency and, most probably, 
the rotational frequency and their multiplets.}

\begin{document}

\maketitle

\section{Introduction}

Binary stars play a special role in astrophysics. An example of their application is the determination of the age
if masses and radii can be determined with a high accuracy.

V380 Cygni (HR 7567, HD 187879, KIC 5385723) is a bright binary \citep[V=5.68,][]{Tycho2000} consisting of two early B-type stars with 
a mass ratio, $M_B/M_A \approx$ 0.6 \citep[e.g.,][]{Batten1962, Popper1998}. The system undergoes shallow 
eclipses (0.12 and 0.09 mag) with the period equal to 12.43 days. In addition, in spectroscopic observations the spectral lines 
of both components are visible \citep{Batten1962}.
Over decades, several papers have been published where masses and radii 
of the components A and B were determined \cite[e.g.,][]{Batten1962, Ramella1980, Popper1998, Guinan2000}. 
The most recent values can be found in \cite{Pavlovski2009}  and  \cite{Tkachenko2014}.
However, the location of the primary component in the HR diagram does not agree with evolutionary tracks computed for determined parameters
\citep{Popper1998, Guinan2000}. The more massive component is overluminous for its mass and, moreover, it is located beyond the main sequence (MS)
in a comparison with models for metallicity close to the solar value and no overshooting from the convective core. The less massive component 
has the parameters which perfectly fit with evolutionary tracks and it is a main sequence object not much evolved from Zero Age Main Sequence (ZAMS). 
\cite{Guinan2000} showed for the first time that in order to catch the primary on MS, the overshooting parameter of at least  $f_{\rm ov}\approx$ 0.06 
was needed if the metallicity was $Z=0.012$.  This results was confirmed also by \cite{Claret2007} and \cite{Tkachenko2014} 
and until now, the evolutionary stage for the primary star has not been resolved.

The V380 Cyg binary was also observed by \textit{Kepler Space Telescope} resulting in eight quarters of accurate 
photometry (Q7, Q9, Q10, Q12, Q11, Q14, Q15, Q16). 
Using the first six quarters of observations \cite{Tkachenko2014} was looking for possible pulsations and announced above 300 frequency peaks.

In Sec.\,2 we describe evolutionary computations. The determination of the age of the system is presented in Sec.\,3. 
In Sect.\,4 we analyze the \textit{Kepler} light curve from all eight quarters. Conclusions end the paper.

\begin{table*}
	\centering
	\caption{The parameters of the two components of the V380 Cyg system determined by Pavlovski et al. (2009) (the first two rows) 
	and Tkachenko et al. (2014) (the next two rows). The subsequent columns contain: the name and HD number, spectral types, 
	the orbital period and brightness at the maximum light, mass, radius, effective temperature, luminosity and the projected rotational velocity.}
	\label{tab:parameters}
	\begin{tabular}{lclccccccc}
		\hline 
		System 	& Star	& SpT	& $P_{\rm orb}$ [d]	& $M$			& $R$							& $T_{\rm{eff}}$ 							& $\log L/L_{\odot}$  			 								& $V_{\rm rot} \sin i$	\\ 
		&			&  	 		& $V_{\rm{max}}$ & [$M_{\rm{\odot}}$] 				& [$R_{\rm{\odot}}$] 					& [K]										&   											&   							[km\,s$^{-1}$] 	\\ 
		\hline 
		
		V380 Cyg	& A	& B1.5II-III	& 12.43	& 13.13$\pm$0.24  			& 16.22$\pm$0.26			& 21750$\pm$280					& 4.723$\pm$0.026							& 98$\pm$4 		\\
		HD 187879	& B	& B2V			& 5.68		& 7.779$\pm$0.095				& 4.060$\pm$0.084			& 21600$\pm$550					& 3.508$\pm$0.048				& 	 										 32$\pm$6 		\vspace{5pt}\\
		
		V380 Cyg	& A	& B1.5II-III	& 12.43	& 11.43$\pm$0.19  			& 15.71$\pm$0.13			& 21700$\pm$0						& 4.691$\pm$0.007							&	 98$\pm$4 	\\
		HD 187879	& B	& B2V			& 5.68		& 7.000$\pm$0.140				& 3.819$\pm$0.048			& 23840$\pm$500					& 3.626$\pm$0.038				& 	 										 32$\pm$6		\vspace{5pt}\\
	\end{tabular} 
\end{table*}

\section{Evolutionary computations}

We assume that V380 Cyg is a detached binary system, so each component evolves separately and there is no interactions
between them.

The evolutionary models were computed using \texttt{MESA} (Modules for Experiments in Stellar Astrophysics) 
evolutionary code  \citep{Paxton2011, Paxton2013, Paxton2015}.
We adopted  the OPAL opacity tables \citep{OPAL1996} and the solar chemical mixture from \cite{Asplund2009}. 

Overshooting from the convective hydrogen core was included according to the prescription proposed by \cite{Herwig2000}, i.e.,: 
\begin{equation}
D_{\rm{ov}}=D_{\rm{conv}} \exp \left( -\frac{2z}{f_{\rm ov}H_{\rm{P}}} \right),
\label{eq:quadratic}
\end{equation}
where $D_{\rm{conv}}$ is the diffusion coefficient derived from the mixing length theory (MLT) inside the convective region, $H_{\rm{P}}$ is the pressure scale height 
at the convective boundary, $z$ is the distance from the convective boundary and $f_{\rm ov}$ is a dimensionless free parameter. 

We used the Ledoux criterion for the convective instability with the value of the mixing-length parameter of  $\alpha_{\rm{MLT}}$=0.5.
This is a usually adopted value for B-type main sequence stars  because of negligible efficiency of the convective transport in thier envelopes.
In convectively unstable regions according to the Schwarzschild criterion but stable according to the Ledoux criterion, a semiconvective mixing 
is included using a formula proposed by \cite{Langer1983, Langer1985}, with the adopted efficiency parameter $\alpha_{\rm{SC}}$=0.01.

\texttt{MESA} code enables to calculate models rotating differentially in the framework of the shellular approximation, which assumes the constant angular velocity $\Omega$, over isobars. 
In our computation rotation and rotational mixing has been included. The effects of mass loss is taken into account according to the prescription by \cite{Vink2001}. 
The effects of the magnetic field were neglected.

\section{Age determination}

To estimate the age of the V380 Cyg system, we used the radius-age diagram at the fixed values of masses.
We adopted the values of masses and radii derived by \cite{Pavlovski2009}, which are listed in Table~\ref{tab:parameters}.
For a comparison, we give also parameters determined by \cite{Tkachenko2014}.

\begin{figure}[h!]
	\includegraphics[width=\columnwidth, clip]{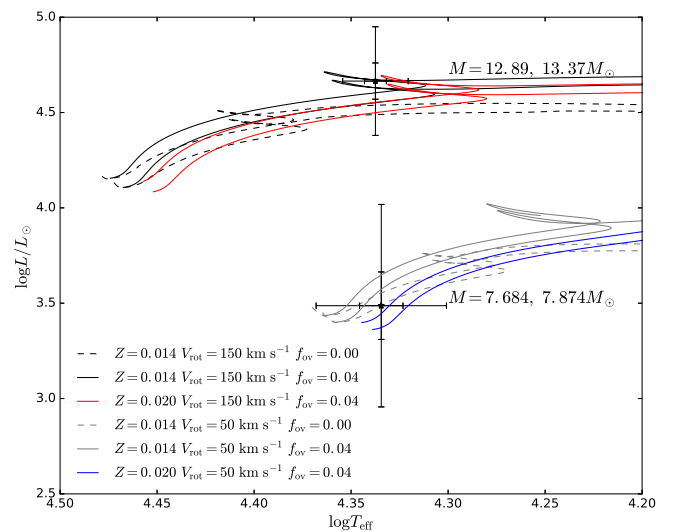}
	\caption{The HR diagram with the position of the components of the V380 Cyg system. There are shown the MESA evolutionary tracks for two values of metallicity and two values of convective overshooting. We assumed the initial rotational velocity 150 km\,s$^{-1}$ for the component A and 50 km\,s$^{-1}$ for the component B. We considered the masses in the range determined from the binary solution.}
	\label{fig:HR}
\end{figure}

In Figure~\ref{fig:HR}, we show the Hertzsprung--Russell diagram with evolutionary tracks computed for the initial hydrogen abundance $X_0=0.70$,
two values of metallicity, $Z=0.014,~0.020$, and two values of the overshooting parameter from the convective core, $f_{\rm ov}=0.00,~0.04$. 
The positions of the stars are marked by crosses with $1\sigma$ and $3\sigma$ 
errors in $\log T_{\rm eff}$ and $\log L/L_{\odot}$. 
We assumed the initial values of rotation \mbox{$V_{\rm{rot}}$=150} km\,s$^{-1}$ for primary and $V_{\rm{rot}}$=50 km\,s$^{-1}$ for secondary in order to fit 
the observed values of rotational velocity within the error (cf. Table.~\ref{tab:parameters}). These values correspond to 22\% and 8\% of the critical velocity 
on ZAMS for the component A and B, respectively. We considered the overshooting parameter from the convective core in the range of $f_{\rm{ov}}\in (0.00,~0.04)$. 
The metallicity of V380 Cyg was estimated by \cite{Prugniel2011} and it is \mbox{$\rm{Fe/H}$=0.05 $\pm$ 0.12}, which corresponds to the abundance of metallicity by mass \mbox{$Z=0.011-0.020$} (assuming $Z_{\odot}=0.0134$). 
Here, we considered two values of the metallicity: $Z=0.014$ as determined by \cite{Nieva2012} for galactic B-type stars
and $Z=0.020$, the highest value determined from observations.

In Figure\,2, we plotted the effective radius, $R_{\rm eff}$, as a function of time, $\log t$, for both components. 
The values of age, $t$, are given in years.
The effective radius is the value of $R$ corrected for the centrifugal force.
The same set of parameters as in Figure~\ref{fig:HR} was adopted.
The age is estimated for each star separately and then a common age is set. As one can see from Figures\,1 and 2,  two hypotheses 
regarding the evolutionary stage of V380 Cyg primary have to be considered. Depending on the adopted parameters, the star can be 
either in post--main sequence phase of evolution or it can be still on the main sequence.

\subsection{Post-main sequence hypothesis}
\label{PMS}

The scenario that the V380 Cyg primary is beyond the main sequence does not require adding overshooting in order to agree the star's position in the HR diagram. 
Keeping the rotational velocity of both components consistent with the observed values, the common age of V380 Cyg depends on the assumed chemical composition.

For $Z=0.014$, the common age of the two components is $\log t_{\rm AB}$=7.11$\pm$0.01 (12.9$\pm$0.3 Myr), 
whereas for $Z=0.02$ $\log t_{\rm AB}$=7.125$\pm$0.015 (13.3$\pm$0.5 Myr).  
The determined ranges of the common age are given in Table~\ref{tab:age}. There are also given: the metallicity, overshooting parameter
and central hydrogen abundance.

\subsection{Main sequence hypothesis}
\label{MS}

In order to catch more massive component of V380 Cyg on the main sequence the large value of the overshooting parameter from the convective core, $f_{\rm{ov}}$=0.04
is required (cf. Figures\,1 and 2).  For $Z=0.014$ we obtained the common age of V380 Cyg of $\log t_{\rm AB}$=7.195$\pm$0.015 (15.7$\pm$0.6  Myr), 
whereas for $Z=0.020$ the age is slightly older, $\log t_{\rm AB}$=7.205$\pm$0.015 (16.0$\pm$0.6 Myr).  

In general, the higher the metallicity the younger the system, 
but because the primary is very close to TAMS the effect is inverse, i.e.,  higher metallicity gives the older age \citep{JDD2018}.
In the case of V380 Cyg the effect of metallicity is very slight because the dependence $R_{\rm eff}(\log t)$  for the primary star is almost vertical (cf. Figure\,2).
For more parameters of the models see Table~\ref{tab:age}.

\vspace{10pt}

\begin{figure}[h!]
	\includegraphics[width=\columnwidth, clip]{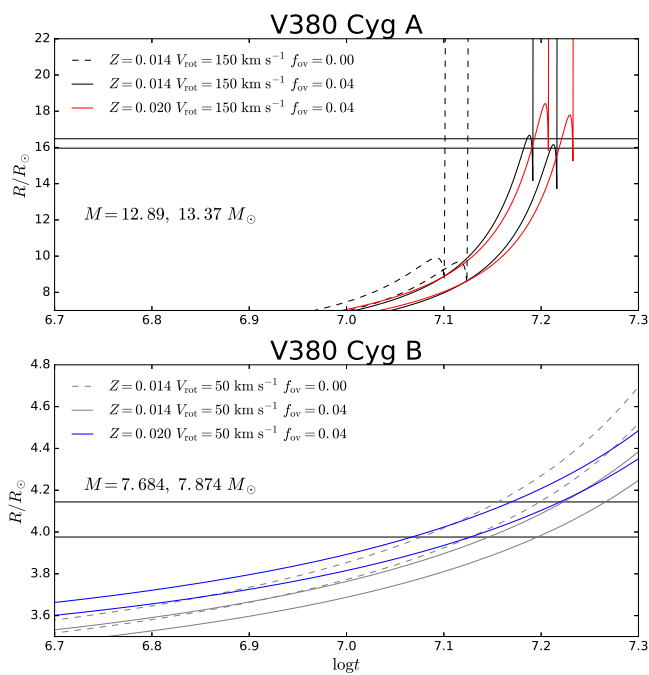}
	\caption{The evolution of the effective radius, $R_{\rm eff}$, for both components of the V380 Cyg system for the masses determined from the binary solution. 
	The values of the age, $t$, are in years. The horizontal lines indicate  the observed range of the radii. The same values of metallicity and convective overshooting 
	were assumed as in the Figure~\ref{fig:HR}.  The ZAMS values of rotation are given in the legend.}
	\label{fig:logt}
\end{figure}

\begin{table*}
	\centering
	\caption{The age range for individual components of V380 Cyg and the common age of the system. The solution was obtained for the two values of metallicity, $Z=0.014$
	 and $Z=0.020$, considering two evolutionary stages: a post--main sequence phase and main sequence phase. Subsequent columns contain:
	 a component, evolutionary stage of each star, metallicity, the overshooting parameter, age of individual component, common age of the system and the central hydrogen abundance. }
	\label{tab:age}
	\begin{tabular}{lcccccccccrrl}
		\hline
		Star& Stage		 & $Z$ 		  & $f_{\rm ov}$ 	& $\log (t/{\rm yr})$			   & $\log (t_{\rm AB}/{\rm yr})$  & $X_{\rm c}$  			\\
		\hline
		
		A	& Post-MS	& 0.020		& 0.0					 & 7.11 -- 7.14		  & 7.11 -- 7.14	  & 0.00				 \\
		B	& MS		   & 			   & 0.0					&7.00 -- 7.15		& 						 & 0.58 -- 0.52	\vspace{5pt}\\
		
		A	& Post-MS	& 0.014		& 0.0					 & 7.10 -- 7.12		  & 7.10 -- 7.12	  & 0.00				\\
		B	& MS		   & 			   & 0.0				    & 7.08 -- 7.20		 & 						  & 0.53 -- 0.47	\vspace{5pt} \\
		\hline
		
		A	& MS		   & 0.020	   & 0.04					& 7.19 -- 7.22	 	 & 7.19 -- 7.22		& 0.07 -- 0.05	\\
		B	& MS		   & 			   & 0.04					& 7.07 -- 7.22		& 						 & 0.60 -- 0.56	\vspace{5pt}\\
		
		A	& MS		   & 0.014		& 0.04					 & 7.18 -- 7.21		  & 7.18 -- 7.21	& 0.04 -- 0.01 \\
		B	& MS		   & 			   & 0.04					 & 7.15 -- 7.27		  & 					  & 0.57 -- 0.52  \vspace{5pt}\\
		
	\end{tabular}
\end{table*}

\section{Kepler photometry}

To obtain the light curve of V380 Cyg, we used all available {\it Kepler} Long Cadence data from eight quarters of observations: 
Q7, Q9, Q10, Q12, Q11, Q14, Q15, Q16. 
The flux from Simple Aperture Photometry (SAP) was extracted from target pixel files \citep[for details see][]{Szewczuk2018}. 
As a result we obtained 34\,051 data points spanned over 928 days.

The eclipsing binary light curve was modelled using the Wilson-Devinney (\texttt{WD})  code \citep[e.g.,][]{WD1971, Wilson2014} in its version of May 22, 2015. 
In this version, the {\it Kepler} passband is included, which enables us to model properly passband-dependent features. Calculations were performed in MODE\,2 dedicated 
for detached binaries in which luminosity of the secondary star is coupled to the temperature. 

We fitted binary model to the all data points simultaneously. The values of fixed and fitted parameters of the model are listed in Table\,\ref{tab:EB}. 
Fixed parameters were taken from \cite{Tkachenko2014}.

\begin{table}
	\centering
	\caption{The fitted and fixed parameters from the \texttt{WD} solution of the \textit{Kepler} light curve of V380 Cyg.}
	\label{tab:EB}
	\begin{tabular}{rrc}
		\hline
		Parameter & Value \\
		\hline
		\textbf{\textit{Fitted parameters:}}  &\\
		Orbital period [d]  & 12.425745106(2) \\
		Longitude of periastron [rad] & 2.07(3) \\
		Change longitude of &    \\
		periastron [rad d$^{-1}$]     &    0.000024(2) \\
		Star A potential  & 4.770(3) \\
		Star B potential  & 11.460(9)\\
		Orbital inclination [\textdegree] &  81.34(3)\\
		$T_{\rm eff}$ of the B star [K]  & 23460(33) \\
		\hline
		
		\textbf{\textit{Fixed parameters:}} &\\
		Primary eclipse  & \\
		Time [HJD] & 2441256.544\\
		Mass ratio & 0.6129 \\
		$T_{\rm eff}$ of the A star [K] & 21700 \\
		Orbital eccentricity  & 0.2224\\
		Orbital semimajor axis [R$_\odot$]  & 59.60 \\

	\end{tabular}
\end{table}

\begin{figure}[h!]
	\includegraphics[width=\columnwidth]{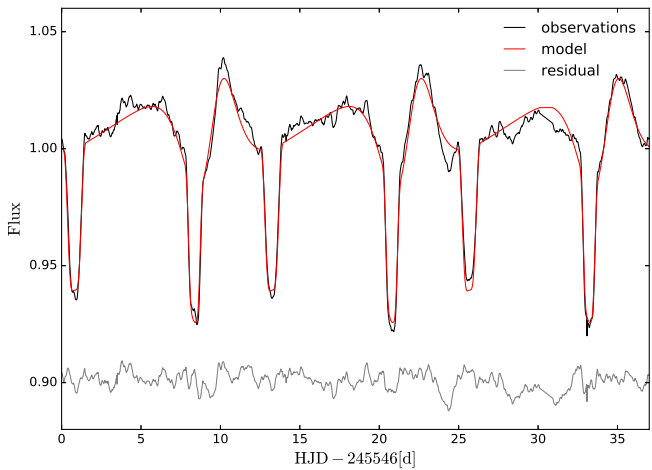}
	\caption{The part of the \textit{Kepler} light curve of the eclipsing binary V380 Cyg covering 37 days. The best model obtained with the \texttt{WD} code is plotted as a red line. 
	The total time span of the data is about 900 days. The bottom grey curves are residuals.}
	\label{fig:model}
\end{figure}

Then, we subtracted modelled eclipsing binary light curve from the observed data (see Figure~\ref{fig:model}) and cut the eclipses. After this operation 
we ended up with 27\,124 data points that were used for searching possible pulsational variability. To this end, we calculated the Fourier amplitude spectrum up
to the Nyquist frequency ($\sim$24.5\, d$^{-1}$) and followed standard pre-whitening procedure.

As a significant signal in the Fourier amplitude spectrum, we treated peaks with signal-to-noise ratio higher than 4 (S/N$>$4). 
Only five frequencies above this threshold were found (see Table\,\ref{tab:freq}), comparing to over 300 peaks found by \cite{Tkachenko2014} 
from six quarters of \textit{Kepler} observations. The reason of such different result is not clear. Partly, this may be due to shorter time series 
they analyzed or possibly results from not accurate subtraction of the modelled binary light curve.

The first frequency, $f_1$, is very close to the rotational frequency ($\nu_{\rm rot}=0.12189$ d$^{-1}$  for $V_{\rm{rot}}$=100 km s$^{-1}$ and 16.22 $R_{\odot}$). 
The frequencies $f_2$ and $f_4$ seem to be its consecutive multiplets ($f_2$=2$f_{\rm{rot}}$ and $f_4$=3$f_{\rm{rot}}$). 
The frequency $f_3$ is very close to the orbital frequency $f_{\rm{orb}}$=0.08045 d$^{-1}$, within the Rayleigh resolution. 
Plausible explanation could be that changes in the brightness with the orbital period occur also outside the eclipses
due to, e.g., tidally-induced deviations from spherical symmetry.  The eccentricity of the system is $e=0.22$ and the shape of the light curve (cf. Figure\,3)
allows us to suppose that we are dealing with a heartbeat star.

The last significant frequency, $f_5$=0.02252 d$^{-1}$, is rather spurious since V380 Cyg is overexposed in the {\it Kepler} observations.

\begin{table}[h!]
	\centering
	\caption{Results of the {\it Kepler} light curve frequency analysis matching the S/N$>$4.0. 
	The following columns contain: frequencies, amplitudes, their errors  and signal-to-noise ratio.}
	\label{tab:freq}
	\begin{tabular}{lcccccccccr}
		\hline
		$f_{\rm i}$	& Freq. [d$^-1$]		& Amplitude	& $\sigma_{\rm \nu}$ 		& $\sigma_{\rm A}$	& S/N \\
		\hline
		$f_1$   & 0.12944  &	 0.001236  &        0.00002   &      33			& 5.59	\\
		$f_2$   & 0.25884  & 	0.001216   &        0.00002   &      33			& 5.77	\\
		$f_3$   & 0.08056  &    0.00996    &        0.00002   &      30 			& 5.10	\\
		$f_4$   & 0.38848  &    0.00952    &        0.00002   &      33			& 5.00	\\
		$f_5$   & 0.02252  &    0.00912    &        0.00002   &      33			& 4.04	\\
		\hline
	\end{tabular}
\end{table}

\section{Conclusions}

We presented the age determination of V380 Cyg, a double-lined eclipsing binary with two massive, early-B type components. 
To this end, we calculated evolutionary models for each component using \texttt{MESA} code and used the mass-radius-age relation.
The estimated age of the V380 Cyg binary depends on the assumed evolutionary stage of the primary component, which can be either 
a main sequence or post-main sequence star.

In order to catch the more massive component on main sequence large overshooting from the convective core is needed ($f_{\rm ov}\approx 0.04$). 
Then, the age of the system depends on the adopted value of the metallicity. We obtained the age of 15.7$\pm$0.6 Myr and 16.0$\pm$0.6 Myr,
for $Z=0.014$ and  $Z=0.020$, respectively.

If the primary star has already entered  the post-MS evolutionary phase, then no overshooting is needed to agree the age of both components.
In this case we derived the age of 12.9$\pm$0.3 Myr and 13.3$\pm$0.5 Myr, for $Z=0.014$ and  $Z=0.020$, respectively.

The effect of metallicity depends on the evolutionary stage. For most of the main sequence evolution, the higher metallicity results in a younger age of the system.
However, the effect is reverse close and after the TAMS and for higher metallicity one gets the older age.

From these studies it is hard to conclude what is the evolutionary stage of the V380 Cyg primary. 
On one hand, the main sequence hypothesis seems more likely due to time scales. On the other hand, it requires large core overshooting 
and for now we are not able to check whether this result has a physical meaning.

Finally, we made an attempt to find pulsational frequencies from the {\it Kepler} light curve. The aim was to get some constraints 
on the evolutionary stage of the primary star from a comparison of pulsational instability with the observed frequency range.
We used Wilson-Devinney code to get an orbital solution from the eight quarters of the {\it Kepler} Long Cadence observations.
After subtracting the modelled light curve and cutting the eclipses, the frequency analysis was performed.

This analysis revealed only 5 significant frequencies, i..e, with the adopted criterion S/N$>$4, comparing to above 300 found by \cite{Tkachenko2014}.
Three our frequencies agree with the values reported by \cite{Tkachenko2014} ($f_1$, $f_2$ and $f_3$). 
The frequency $f_1=0.12944$\,d$^{-1}$ is most probably the rotational frequency. The other two seem to be its multiplets.  
Despite removing the eclipses, the orbital frequency is still present. This can indicate, e.g., distortion of stars as a result of tidal interaction.
Given that eccentricity is not small ($e=0.22$), it is quite likely that V380 Cyg could be a heartbeat binary. 
More studies are needed to confirm this suggestion. 

\section*{Acknowledgments}
This work was financially supported by the Polish National Science Centre grant 2015/17/B/ST9/02082.

\bibliographystyle{phostproc}
\bibliography{V380_Cyg_v6.bib}

\end{document}